\documentclass[useAMS,usenatbib]{mn2e}

\usepackage{graphicx}

\title[Growth of supermassive black holes]{The growth of supermassive black holes in pseudo-bulges,
classical bulges and elliptical galaxies}
\author[D. A. Gadotti \& G. Kauffmann]{Dimitri A. Gadotti\thanks{E-mail: dimitri@mpa-garching.mpg.de}
and Guinevere Kauffmann\\
Max-Planck-Institut f\"ur Astrophysik, Karl-Schwarzschild-Str. 1, D-85748
Garching bei M\"unchen, Germany}
\begin{document}


\pagerange{\pageref{firstpage}--\pageref{lastpage}} \pubyear{2008}

\maketitle

\label{firstpage}

\begin{abstract}
Using results from structural analysis of a sample of nearly 1000 local galaxies from the Sloan
Digital Sky Survey, we estimate how the mass in central black holes is distributed amongst
elliptical galaxies, classical bulges and pseudo-bulges, and
investigate the relation between their stellar masses and central stellar velocity dispersion $\sigma$.
Assuming a single relation between elliptical galaxy/bulge mass, $M_{\rm{Bulge}}$, and central
black hole mass, $M_{\rm{BH}}$, we find that $55^{+8}_{-4}$ per cent of the mass in black holes in the local universe
is in the centres of elliptical galaxies, $41^{+4}_{-2}$ per cent in classical bulges and $4^{+0.9}_{-0.4}$ per cent in pseudo-bulges.
We find that ellipticals, classical bulges and pseudo-bulges follow different relations between
their stellar masses and $\sigma$, and the most significant offset occurs
for pseudo-bulges in barred galaxies. This structural dissimilarity leads to discrepant black hole
masses if single $M_{\rm{BH}}-M_{\rm{Bulge}}$ and $M_{\rm{BH}}-\sigma$ relations are used.
Adopting relations from
the literature, we find that the $M_{\rm{BH}}-\sigma$ relation yields an estimate of the total mass density
in black holes that is roughly 55 per cent larger than if the $M_{\rm{BH}}-M_{\rm{Bulge}}$ relation is used.
\end{abstract}

\begin{keywords}
galaxies: bulges -- galaxies: evolution -- galaxies: formation -- galaxies: fundamental parameters --
galaxies: kinematics and dynamics -- galaxies: photometry
\end{keywords}

\section{Introduction}
\label{sec:intro}

In the past 10 years or so, there has been mounting evidence that massive galaxies host supermassive
black holes in their centres, and that the mass of the black hole correlates with other galaxy
properties, particularly luminosity, stellar mass (or bulge luminosity and
stellar mass in the case of disc galaxies) and central velocity dispersion
\citep[see e.g.][and references therein]{MagTreRic98,GebBenBow00}. Such relations are being discussed
and revised, with several important details being disclosed
\citep{TreGebBen02,HarRix04,FerFor05,GraDri07,Ber07,BerSheTun07,
TunBerHyd07,LauTreRic07}, and a consensus emerges that they reveal
a connected growth of black holes and their host galaxies or bulges, e.g. via mechanisms of feedback
\citep[e.g.][]{WyiLoe03,DiMSprHer05,YouHopCox08}. With growing evidence that galaxy bulges are
not a single, homogeneous class, but in fact comprise classical bulges and pseudo-bulges, with
different formation histories \citep[see e.g.][and references therein -- hereafter Paper I]{Gad09},
a new ingredient is added to this investigation. Do black holes in elliptical galaxies, classical
bulges and pseudo-bulges follow similar relations between their masses and host galaxy/bulge mass
or velocity dispersion? \citet{Hu08} suggests that the relation between black hole mass
$M_{\rm{BH}}$ and velocity dispersion $\sigma$ is different for classical bulges (including there
elliptical galaxies) and pseudo-bulges.

To address this issue, one would ideally have secure, direct measurements of $M_{\rm{BH}}$ for a statistically
significant sample of galaxies, which is currently not available. In this Paper, we approach the question by
using measurements of the stellar mass in elliptical galaxies and bulges, obtained in Paper I, for
a sample of nearly 1000 systems from the Sloan Digital Sky Survey (SDSS). We combine these results with
$\sigma$ measurements from SDSS, and investigate how the stellar mass of the bulge
relates to $\sigma$ in ellipticals, classical
bulges and pseudo-bulges. By assuming that we can infer $M_{\rm{BH}}$ from bulge stellar masses using
a single relation, we indirectly assess the $M_{\rm{BH}}-\sigma$ relation for these systems.

In the next section, we briefly recall how the bulge stellar mass measurements were done, as well as how ellipticals,
classical bulges and pseudo-bulges were defined, and address our use of SDSS $\sigma$ measurements. In Sect.
\ref{sec:results}, we show the results, which are discussed in Sect. \ref{sec:dis}.

\section{Data}
\label{sec:pap1}

In Paper I, we have performed careful and detailed image fitting of the galaxies in the sample
in the $g$, $r$ and $i$ bands, including up to three components in the models, namely bulge, disc
and bar. This allowed a reliable determination of the bulge luminosity.
The presence of these components was assessed by individual inspection of images,
surface brightness profiles and isophotal maps. A galaxy is defined as elliptical if it does not
show any signature of a disc,
in which case it was fitted with a single ``bulge'' component. Classical bulges
and pseudo-bulges are separated using the \citet{Kor77} relation, where pseudo-bulges can be identified
as outliers in an objective fashion.

Since we have done multi-band decompositions, we were able to estimate the $g-i$ integrated
colour of each component separately. Using the relation between $g-i$ and the stellar mass-to-light
ratio in the $i$-band from \citet{KauHecBud07}, we have all parameters necessary to accurately calculate
the stellar masses of the ellipticals and bulges in the sample.

The sample was designed to be concomitantly suitable for structural analysis based on
image decomposition and a fair representation of the galaxy population in the local universe. It was
drawn from all objects spectroscopically classified as galaxies in the SDSS Data Release Two (DR2)
at redshifts $0.02\leq z\leq0.07$, and with stellar masses larger than $10^{10}~{\rm M}_\odot$.
At this stage, we have a volume-limited sample of {\em massive} galaxies, i.e. a sample which includes
all galaxies more massive than $10^{10}~{\rm M}_\odot$ in the volume defined by our redshift
cuts and the DR2 footprint.
In order to produce reliable decompositions, and avoid dust and
projection effects, we have applied another important selection criterion to produce the final sample:
it contains only galaxies close
to face-on, i.e. with an axial ratio $b/a\geq0.9$, where $a$ and $b$ are, respectively, the
semi-major and semi-minor axes of the galaxy at the 25 $g$-band mag arcsec$^{-2}$ isophote.
We have found that this introduces a selection effect, in the sense that the probability
of selecting an elliptical galaxy is a factor of 1.3 larger than that of selecting a
disc galaxy. Taking this selection effect into account, we found that the final sample is
representative of the local population of massive galaxies. This was done by comparing the distributions
of several main galaxy properties, such as absolute magnitude, D$_n$(4000) and concentration,
in the volume-limited and final samples, and verifying that these distributions are similar.
The reader is referred to Paper I for a detailed account of the sample selection and
image decomposition.

We have used $\sigma$ measurements from SDSS Data Release 6 (DR6), since estimates from previous
releases can be overestimated in the case of low mass galaxies (see discussion in Paper I,
Sect. 4.4). Since the spectral resolution of SDSS spectra is limited, one should avoid spectra
with signal-to-noise ratio below 10, or with warning flags. In the sample, only one galaxy
does not comply with the former criterion, and only 8 do not comply with the latter.
However, $\sigma$ is available only for spectra which are identified as being from early-type
galaxies through Principal Component Analysis \citep[see][]{BerSheAnn03a,ConSza99}. This
is done in order to exclude spectra with e.g. strong emission lines, as these features can lead
to wrong estimates of $\sigma$. In our sample, this criterion tends to exclude bulges with more
significant star formation. In fact, we have $\sigma$ estimates for only 30 per cent of the pseudo-bulges
in our sample. The corresponding fractions for classical bulges and ellipticals are, respectively,
60 and 76 per cent. Nevertheless, we have verified that $\sigma$ is generally available for galaxies
with values of D$_n$(4000) greater than about 1.4, and thus only those bulges with the strongest star
formation are not represented here (see Paper I, Fig. 9).
The pseudo-bulges in our sample have relatively low $\sigma$ values,
close to the SDSS instrumental resolution (70 km/s). In fact, 24 of the 61
pseudo-bulges for which $\sigma$ is measured have $\sigma$ below 70 km/s. We decided
to keep these measurements for reasons which will be clear below. However, they should be
considered as upper limits. We applied aperture corrections to the SDSS measurements,
using the prescription by \citet{JorFraKja95}, obtaining $\sigma$
at 1/8 of the bulge effective radius $\sigma_{e/8}$. This prescription is based on
measurements for bulge-dominated galaxies, and it is unclear if it is also valid for disc-dominated
galaxies. As we will discuss below, the corrections applied are small, and do not affect our results
significantly.

\section{Results}
\label{sec:results}

If the ellipticals and bulges in our sample host supermassive black holes in their centres,
then we can estimate the black hole masses, using the relation between elliptical galaxy/bulge
mass and black hole mass from \citet[a significant update of the results in
\citealt{MagTreRic98}]{HarRix04} and our elliptical
galaxy/bulge mass measurements. Furthermore, we can also see how the mass
in black holes is distributed amongst the different galaxies.
We have thus computed the black hole mass in each elliptical galaxy and bulge in our sample
directly from the \citet{HarRix04} relation. Adding up the masses of all black holes
in our sample we obtain a total black hole mass, with which we can compute what
fraction of this total mass is in ellipticals and what fraction is in bulges.
We find that 55 per cent
of the mass in supermassive black holes in the local universe is in the centres of elliptical
galaxies, 41 per cent in classical bulges and 4 per cent in pseudo-bulges, after accounting for
the selection effect introduced by the axial ratio cut, discussed in detail in Paper I
(see also Sect. \ref{sec:pap1}). The uncertainty in these fractions from
Poisson statistics only is between 1 and 2 percentage points. However, the
intrinsic scatter around the \citet{HarRix04} relation is 0.33 dex \citep[see][]{TunBerHyd07},
which should dominate the uncertainties in these fractions we report.
To find the uncertainties arising from the scatter in the \citet{HarRix04} relation,
we first calculated the uncertainty in each black hole mass using a constant value of
0.33 dex converted to linear units. For each black hole mass, the upper and lower limit
uncertainties (respectively $\Delta_{\rm up}$ and $\Delta_{\rm low}$) are thus

\begin{equation}
\begin{array}{l}
\displaystyle\Delta_{\rm up}=10^{(\log M_{\rm{BH}}+0.33)}-M_{\rm{BH}}\\
\displaystyle\Delta_{\rm low}=M_{\rm{BH}}-10^{(\log M_{\rm{BH}}-0.33)},
\end{array}
\end{equation}

\noindent where $M_{\rm{BH}}$ is the result from the \citet{HarRix04} relation
in linear units. We then computed the uncertainties in the total black hole mass
in ellipticals, $\sigma_{\rm ell,up}$ and $\sigma_{\rm ell,low}$, through error propagation:

\begin{equation}
\begin{array}{l}
\displaystyle\sigma_{\rm ell,up}=\sqrt{\sum \Delta^2_{\rm up}}\\
\displaystyle\sigma_{\rm ell,low}=\sqrt{\sum \Delta^2_{\rm low}},
\end{array}
\end{equation}

\noindent where the sums concern elliptical galaxies only. Similar equations were used
to compute the uncertainties in the total black hole mass in bulges and
in all galaxies. Finally, to find the upper and lower limit uncertainties in the
fraction of the total black hole mass that are in ellipticals, $\sigma_{\rm f,ell,up}$
and $\sigma_{\rm f,ell,low}$, respectively, we again used error propagation formulae:

\begin{equation}
\begin{array}{l}
\displaystyle\left(\frac{\sigma_{\rm f,ell,up}}{f_{\rm ell}}\right)^2=\left(\frac{\sigma_{\rm ell,up}}{M_{\rm BH,ell}}\right)^2+\left(\frac{\sigma_{\rm tot,up}}{M_{\rm BH,tot}}\right)^2\\
\displaystyle\left(\frac{\sigma_{\rm f,ell,low}}{f_{\rm ell}}\right)^2=\left(\frac{\sigma_{\rm ell,low}}{M_{\rm BH,ell}}\right)^2+\left(\frac{\sigma_{\rm tot,low}}{M_{\rm BH,tot}}\right)^2,
\end{array}
\end{equation}

\noindent where $f_{\rm ell}$ is the fraction of the total black hole mass in ellipticals,
$M_{\rm BH,ell}$ is the total black hole mass in ellipticals, $\sigma_{\rm tot,up}$
and $\sigma_{\rm tot,low}$ are respectively the upper and lower limit uncertainties in the total
black hole mass, and $M_{\rm BH,tot}$ is the total black hole mass in all galaxies in our sample.
Note that Eqs. (3) do not take into account the covariance term between $M_{\rm BH,ell}$ and
$M_{\rm BH,tot}$. The effect of the covariance term is to lower the estimated uncertainties.
Similar equations were used to calculate the uncertainties in the fractions of the total
black hole mass that are in classical and pseudo-bulges. We can now quote the fractions we
find with the estimated uncertainties arising from the intrinsic scatter in the \citet{HarRix04} relation:
$55^{+8}_{-4}$ per cent of the mass in black holes is in elliptical
galaxies, $41^{+4}_{-2}$ per cent in classical bulges and $4^{+0.9}_{-0.4}$ per cent in pseudo-bulges.

We note that there is no particular reason why black hole masses obtained using
a relation between bulge mass, $M_{\rm{Bulge}}$, and black hole mass are
more correct than those obtained through an $M_{\rm{BH}}-\sigma$ relation.
We could have used as well an $M_{\rm{BH}}-\sigma$ relation from the literature
to obtain black hole masses. We have used the \citet{HarRix04} $M_{\rm{BH}}-M_{\rm{Bulge}}$
relation to obtain black hole masses for all galaxies simply because we do not
have measurements of $\sigma$ for all our galaxies, and thus this does not imply that the
$M_{\rm{BH}}-M_{\rm{Bulge}}$ relation is to be preferred over the $M_{\rm{BH}}-\sigma$ relation.
The consequences of this choice are discussed below when necessary. \citet{LauTreRic07} and
\citet{TunBerHyd07} provide extensive discussion on the consequences of using different
relations to infer black hole masses \citep[see also][]{Ber07}.

\begin{figure*}
   \centering
   \includegraphics[keepaspectratio=true,width=7cm,clip=true]{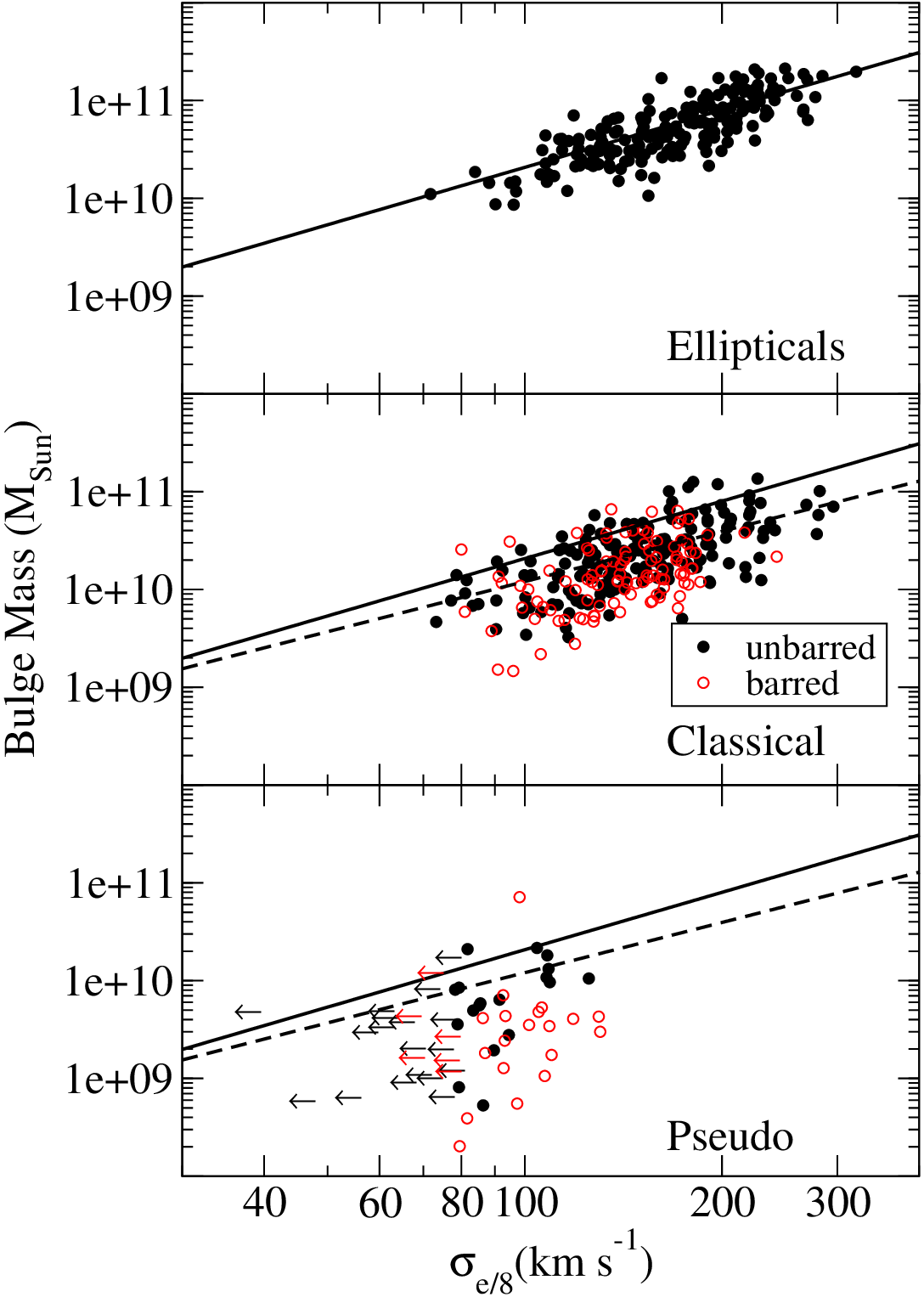}
   \includegraphics[keepaspectratio=true,width=7cm,clip=true]{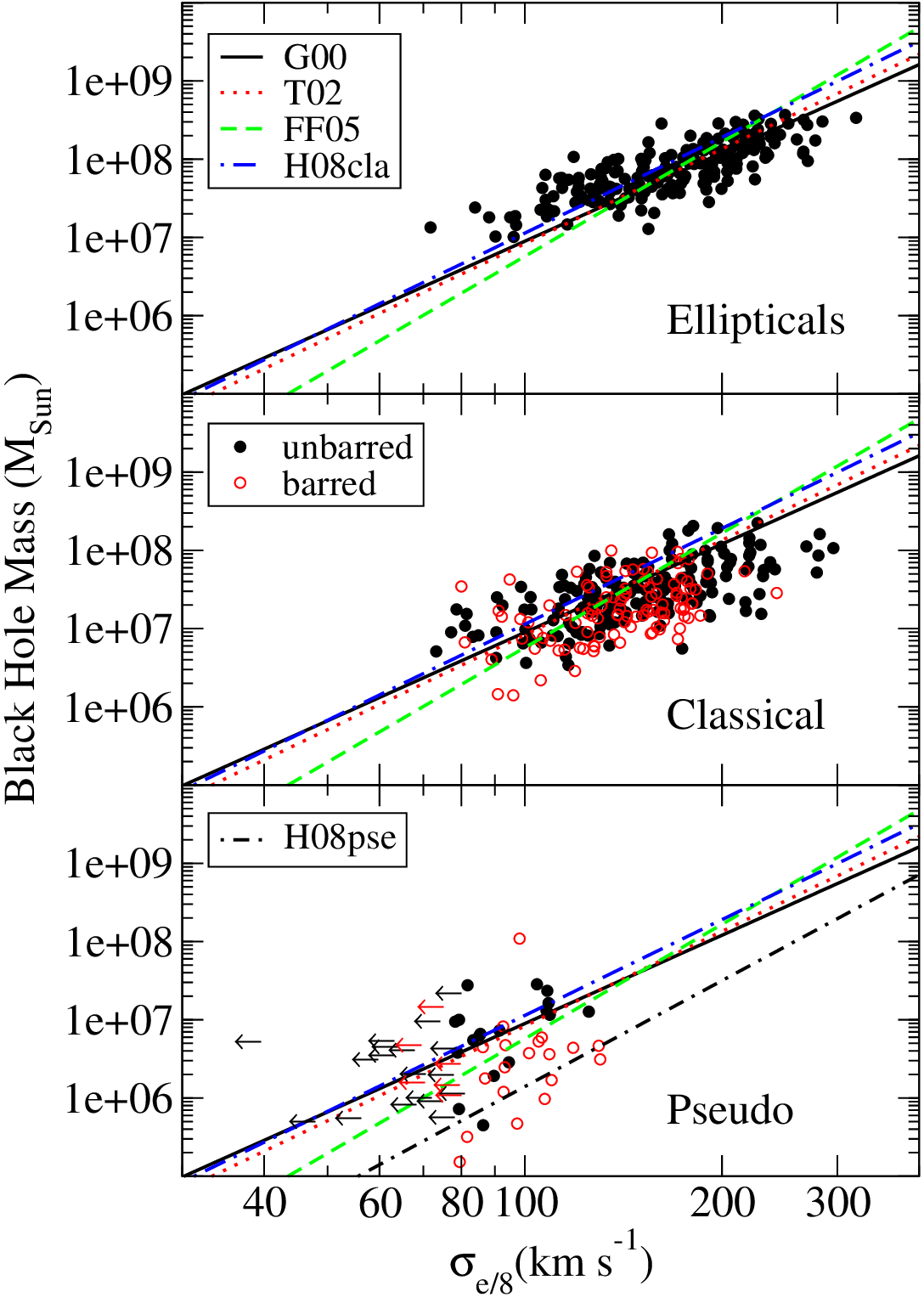}
   \caption{Left: bulge mass plotted against velocity dispersion for elliptical galaxies, classical bulges and
pseudo-bulges, as indicated. The solid line is a fit to the ellipticals, while the dashed line is a fit
to the classical bulges. Right: black hole mass [obtained from the $M_{\rm{BH}}-M_{\rm{Bulge}}$ relation
in \citet{HarRix04}] against velocity dispersion. The lines are relations obtained
from the literature. Arrows indicate those galaxies with velocity dispersion measurements below 70 km/s. Their
$\sigma_{e/8}$ values should be considered as upper limits. [G00: \citet{GebBenBow00};
T02: \citet{TreGebBen02}; FF05: \citet{FerFor05}; H08cla: \citet[classical bulges]{Hu08};
H08pse: \citet[pseudo-bulges]{Hu08}.]}
   \label{fig:smbh}
\end{figure*}

Using the $\sigma_{e/8}$ values, we can also check how bulge mass and black hole mass relate
to velocity dispersion in ellipticals, classical bulges and pseudo-bulges.
This is done in Fig. \ref{fig:smbh}.
One sees that elliptical galaxies follow a well-defined relation between their stellar masses and $\sigma_{e/8}$,
as expected from the \citet{FabJac76} relation. Classical bulges deviate slightly from this relation and follow
a somewhat offset line, with lower masses for the same velocity dispersion.
Pseudo-bulges tend to fall far off
the ellipticals' relation, being on average much less massive than one would expect from this relation.
Note also that one cannot see a {\em clear} relation between bulge mass and $\sigma_{e/8}$
for pseudo-bulges alone.
The fits to the data shown in the left panels of Fig. \ref{fig:smbh} were obtained by minimising $\chi^2$ as
in Eq. (3) of \citet{TreGebBen02}, i.e. weighting every point by the inverse of
its measurement uncertainties.
We use the uncertainties in $\sigma_{e/8}$ as provided by the SDSS. For the uncertainties in bulge mass we
use 0.1 dex, i.e. the same fractional uncertainty in all bulge mass estimates. Such procedure
has also been used by \citet{GebBenBow00} in fitting the $M_{\rm{BH}}-\sigma$ relation.
We have chosen the value of 0.1 dex because this is the typical uncertainty in the estimates
of galaxy masses when one uses colours to derive mass-to-light ratios \citep{KauHecWhi03,KauHecBud07},
as we have done. Note, however, that while this value is a safe estimate for the uncertainty in the masses of
the ellipticals, it is only a lower limit in the case of bulges, as the uncertainty in bulge luminosity
from the image decomposition of disc galaxies is not taken into account (the current version of the
code used to perform the decompositions in Paper I does not provide such estimate).
Nevertheless, we have verified that there is no substantial change in the fits obtained even if no
weighting is included, and the difference in slope obtained in the relations for ellipticals
and classical bulges is statistically significant at $\approx95$ per cent confidence level. Furthermore,
2D Kolmogorov-Smirnov tests indicate that the distributions of bulge mass and $\sigma$ are different
for ellipticals, classical bulges and pseudo-bulges at $\approx99$ per cent confidence level.
These results might be not too surprising, considering that structural differences between pseudo-bulges,
classical bulges and ellipticals are found in Paper I. These differences can
have consequences on the $M_{\rm{BH}}-\sigma$ relation, since black hole mass is correlated with
bulge mass. This is explicitly shown
in the right panels of Fig. \ref{fig:smbh}. The $M_{\rm{BH}}-\sigma$ relation we find for
ellipticals is generally well described by relations found with real black hole mass measurements
\citep{GebBenBow00,TreGebBen02,FerFor05,Hu08}, although our ellipticals seem to follow a somewhat shallower
relation. Evidently, black holes in classical
bulges follow a slightly offset line, while those in pseudo-bulges are, on average, significantly detached
from the ellipticals' $M_{\rm{BH}}-\sigma$ relation.

Despite the small $\sigma$ aperture corrections,
it is legitimate to be concerned about the fact that the extrapolation from $\sigma$ to $\sigma_{e/8}$ is relatively
more significant to pseudo-bulges than to classical bulges and ellipticals. In fact, the mean relative difference
$(\sigma_{e/8}-\sigma)/\sigma$ is 12 per cent for pseudo-bulges, 9 per cent for classical bulges and 5 per cent
for ellipticals. However, since these corrections follow a power law, the results
in Fig. \ref{fig:smbh} do not depend on whether the corrections applied correspond to $\sigma$ at 1/8
of the bulge effective radius or at any other fraction of it, as the difference in such corrections produces only
a constant shift. For instance, we have verified that exactly the same offsets are seen
if we use $\sigma$ at the effective radius, which involves even smaller corrections, instead of $\sigma_{e/8}$.
Furthermore, these results are essentially unchanged even if no aperture correction is applied.

Figure \ref{fig:smbh} thus shows that we find that the $M_{\rm{Bulge}}-\sigma$ relation of classical bulges is
flatter than that of ellipticals. Furthermore, the $M_{\rm{BH}}-\sigma$ relations we find for ellipticals and
classical bulges are also flatter than the relations found in the literature using direct black hole mass
measurements. One should thus verify that this flattening is not caused by our selection effects. In fact,
because the scatter around the $M_{\rm{Bulge}}-\sigma$ relation is larger at the
low mass end, a cut in mass could in principle produce such flattening.
Furthermore, because the uncertainties in $M_{\rm{Bulge}}$ are presumably larger than those in $\sigma$,
$\langle M_{\rm{Bulge}}\vert \sigma\rangle$ could be more affected by such bias than
$\langle \sigma\vert M_{\rm{Bulge}}\rangle$. However, we stress that our cut in mass concerns galaxy
mass, not bulge mass. Most galaxies at the low mass end have significant disc components, and thus
the $M_{\rm{Bulge}}-\sigma$ relations we find are not significantly biased by our mass cut.
In fact, the range in bulge mass in our sample goes as low as more than an order
of magnitude below our mass cut, as do the sample of galaxies with direct $M_{\rm{BH}}$
measurements. Thus our mass cut should not have an important effect in producing the flattening
of the relations we find here.

These findings thus indicate that ellipticals, classical bulges and pseudo-bulges can not
follow a single $M_{\rm{BH}}-M_{\rm{Bulge}}$ relation {\em and} a single $M_{\rm{BH}}-\sigma$ relation.
This conclusion follows directly from the fact that these systems have different
$M_{\rm{Bulge}}-\sigma$ relations. Therefore, it does not depend on whether the
\citet{HarRix04} relation we use here correctly predicts black hole masses.
In order to precisely determine such relations, with direct black hole
mass measurements, one should thus look carefully at the different stellar systems for which such measurements
are available. These differences could partially account for the discordant relations found in the literature
\citep[see in particular discussion in][]{TreGebBen02}. \citet{BerSheTun07} have recently raised and
discussed the fact that $M_{\rm{BH}}$ derived from $\sigma$ is inconsistent with $M_{\rm{BH}}$ derived from
bulge luminosity or mass, in a sample of early-type galaxies from the SDSS. They have also discussed
the importance of the relation between $\sigma$ and bulge luminosity or mass in this regard. We briefly
discuss their results, in connection with our results and others in the recent literature, in
Sect. \ref{sec:dis2}.

Figure \ref{fig:smbh} shows that barred galaxies, particularly with pseudo-bulges, have bulges with lower
masses, at fixed velocity dispersion, on average, than their unbarred counterparts. This is in agreement with
the results from the fundamental plane in Paper I (see Fig. 16).
Indeed, the offset from the $M_{\rm{Bulge}}-\sigma$
relation for pseudo-bulges is caused mostly by barred galaxies. Furthermore, the $M_{\rm{BH}}-\sigma$
relation found by \citet{Hu08} for pseudo-bulges originates {\em almost exclusively} from barred
galaxies. In fact, one sees in the bottom right panel of Fig. \ref{fig:smbh} that his relation describes
reasonably well the pseudo-bulges in barred galaxies in our sample.
This also agrees with the results in \citet{Gra08}.
It is thus interesting to confirm with higher resolution data whether the deviation of pseudo-bulges from the
$M_{\rm{Bulge}}-\sigma$ and $M_{\rm{BH}}-\sigma$ relations occurs regardless of its host galaxy being
barred or unbarred, or if the presence of a bar is a necessary condition, as our results suggest.

It thus seems that studies on black hole demographics
\citep[e.g.][]{YuTre02,WyiLoe03,MarRisGil04,ShaSalGra04,BenDzaFre07} might have different results depending
on whether black hole masses are obtained using an $M_{\rm{BH}}-\sigma$ relation
or an $M_{\rm{BH}}-M_{\rm{Bulge}}$ relation. This comes not only from the possibility of different relations
for ellipticals and bulges, but also from the fact that we find a flatter relation between $M_{BH}$ and
$\sigma$, using the \citet{HarRix04} $M_{\rm{BH}}-M_{\rm{Bulge}}$ relation, than published $M_{\rm{BH}}-\sigma$
relations. This could affect both the total black hole mass density and the black hole mass
distribution. To quantitatively assess how strong such effects can be,
we have recalculated the black hole masses of the galaxies in our sample using the $M_{\rm{BH}}-\sigma$ relation
in \citet{TreGebBen02}. The total black hole mass density using this $M_{\rm{BH}}-\sigma$ relation is $\approx$
70 per cent\footnote{To make this assessment, we use the velocity dispersion measurements from releases prior
to DR6, since most of the studies mentioned used these estimates. If we use the DR6 estimates then the
corresponding difference falls to roughly 40 per cent.}
higher than that using the \citet{HarRix04} $M_{\rm{BH}}-M_{\rm{Bulge}}$
relation, if one does not take into account the intrinsic scatter in these relations.
Interestingly, this is mostly a result
from the different $M_{\rm{BH}}$ estimates in classical bulges. Although black hole masses in pseudo-bulges are by
far more severely discrepant, their contribution to the overall difference in the total black hole mass is small,
due to their small masses, and the fact that the scatter around the $M_{\rm{BH}}-\sigma$ relation in our sample
(see Fig. \ref{fig:smbh}) roughly cancels this effect out. Such scatter
also contributes to reduce the total black hole mass discrepancy in the case of classical bulges and ellipticals.
This scatter comes exclusively from the measurements of $M_{\rm{Bulge}}$ [from which we obtain $M_{\rm{BH}}$ through the
\citet{HarRix04} relation] and $\sigma$ [from which we obtain $M_{\rm{BH}}$ through the \citet{TreGebBen02}
relation]. However, \citet{YuTre02} show that the intrinsic scatter in these relations increases the estimated total
black hole mass density by a factor

\begin{equation}
\exp\left[\frac{1}{2}(\Delta_{\log M_{\rm{BH}}}\ln10)^2\right],
\end{equation}

\noindent where $\Delta_{\log M_{\rm{BH}}}$ is the intrinsic scatter in black hole mass, given $M_{\rm{Bulge}}$
[in the case of the \citet{HarRix04} relation], or given $\sigma$ [in the case of the
\citet{TreGebBen02} relation]. Since the intrinsic scatter in the \citet{HarRix04} relation is 0.33 dex, which
is larger than that in the \citet{TreGebBen02} relation, which is 0.22 dex \citep[see][]{TunBerHyd07}, the
net effect of intrinsic scatter
is to reduce the discrepancy we find in the total black hole mass density using both relations.
It turns out that, taking into account the intrinsic scatter in both relations, the discrepancy falls
$\approx$ 15 percentage points, i.e. to roughly 55 per cent.

\begin{figure}
   \centering
   \includegraphics[keepaspectratio=true,width=7cm,clip=true]{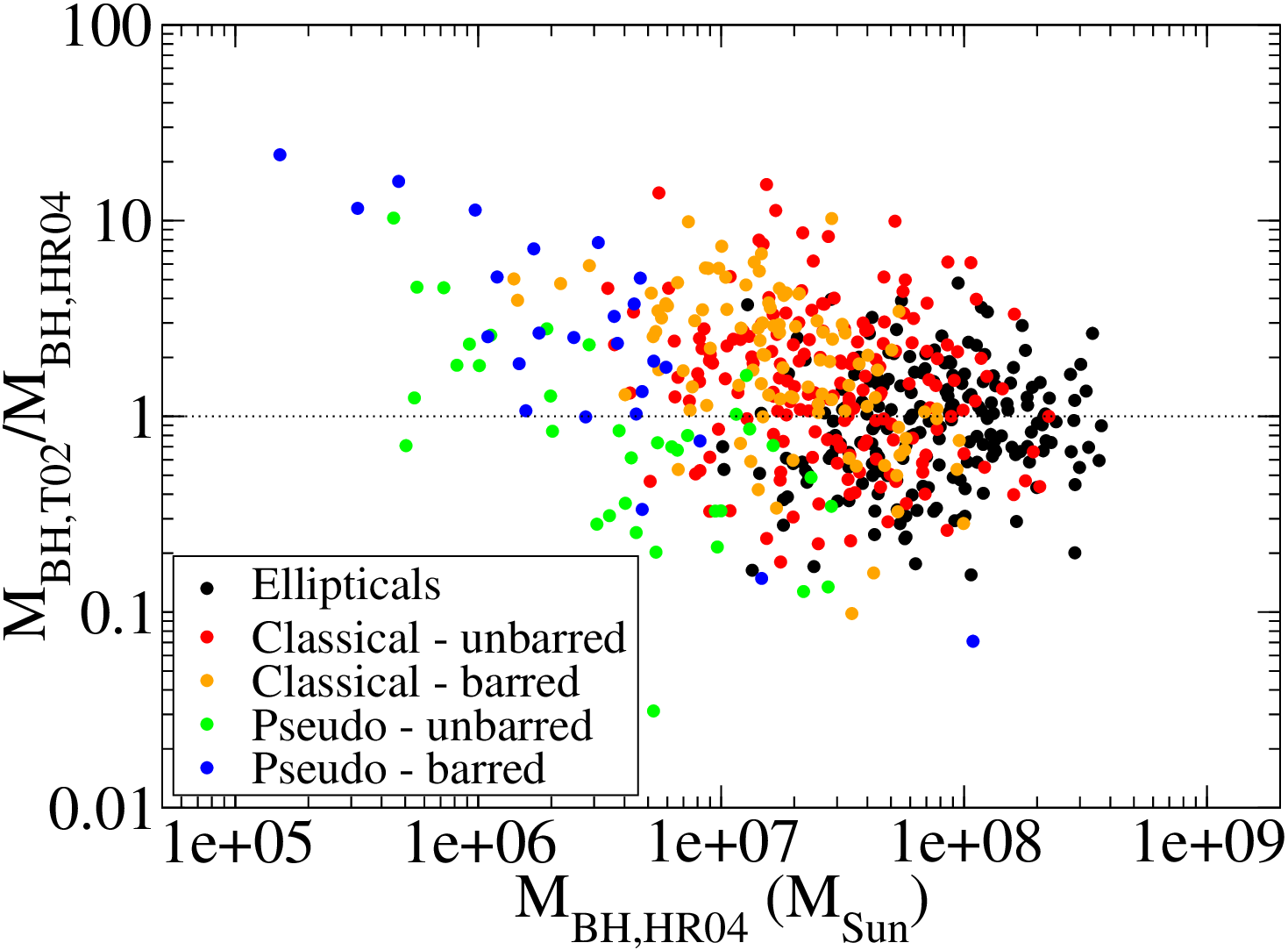}
   \includegraphics[keepaspectratio=true,width=7cm,clip=true]{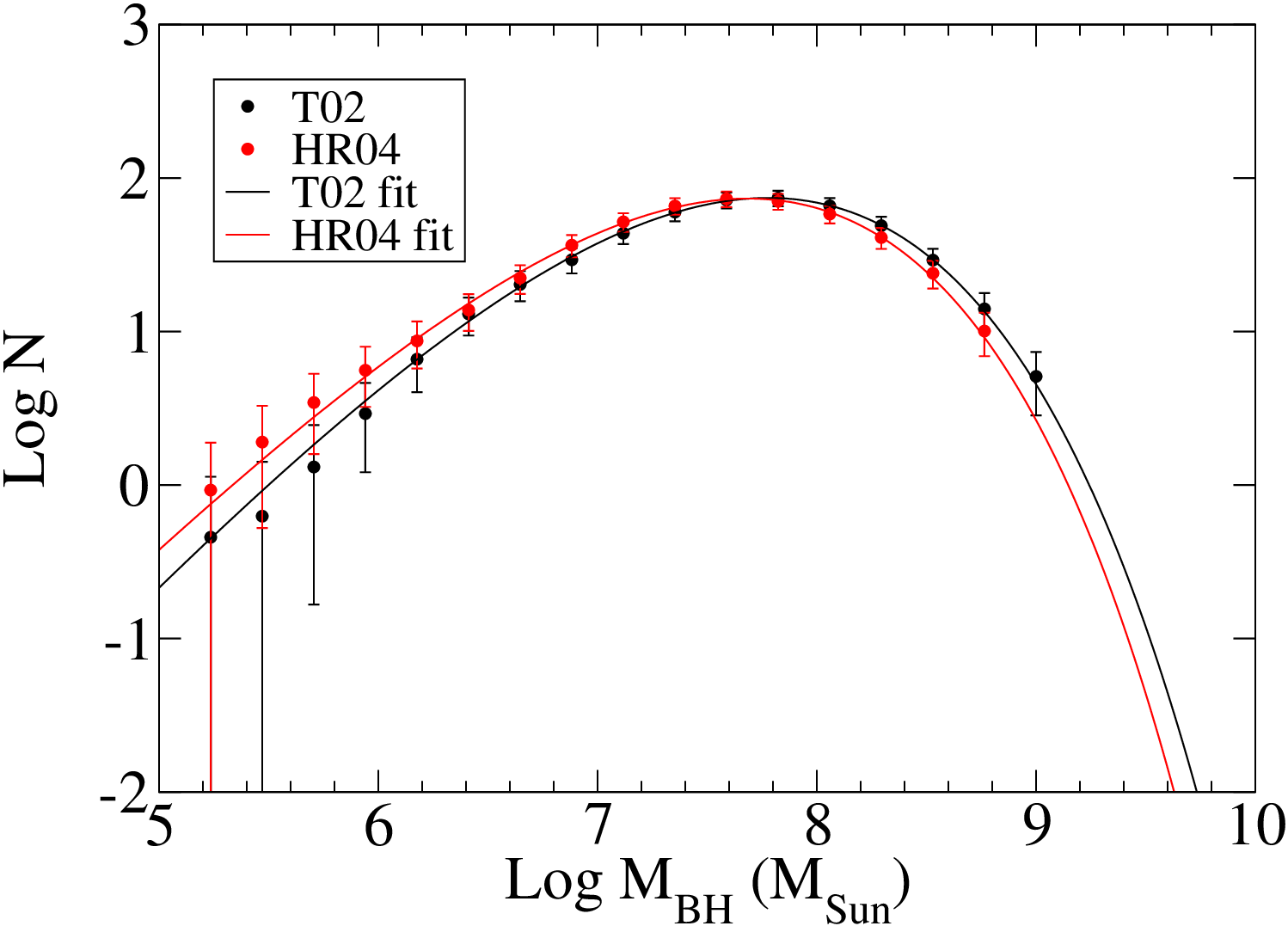}
   \caption{Top: ratio between $M_{\rm{BH}}$ from the $M_{\rm{BH}}-\sigma$ relation in
\citet{TreGebBen02} and $M_{\rm{BH}}$ from the $M_{\rm{BH}}-M_{\rm{Bulge}}$ relation in \citet{HarRix04}
plotted against the latter. Bottom: distributions of $M_{\rm{BH}}$ from the two relations, as indicated.
Error bars are $\sqrt N$.}
   \label{fig:deltabh}
\end{figure}

In Fig. \ref{fig:deltabh}, we explore how the black hole mass distribution varies according to the relation
used to obtain $M_{\rm{BH}}$. The top panel shows explicitly that the difference between the two estimates
is on average negligible at higher masses, increases towards lower masses, and is typically a factor of a few.
The bottom panel shows the corresponding black hole mass distributions.
Again, it is important to take into account the intrinsic scatter in the \citet{HarRix04} and \citet{TreGebBen02}
relations. To do that, we have convolved the distributions obtained directly from our $M_{\rm{BH}}$ estimates
with normal distributions with the appropriate scatter, i.e. 0.33 dex in the case of the \citet{HarRix04}
relation, and 0.22 dex in the case of the \citet{TreGebBen02} relation. The resulting distributions are
shown as data points.
Also shown are fits to these data, using the same fitting function as in
\citet[][their Eq. 4]{ShaSalGra04}. Uncertainties were calculated as $\sqrt N$ (where $N$
is the number of measurements in each bin) and used to weight each data point when determining the fits.
As expected, the distribution obtained using the
\citet{TreGebBen02} $M_{\rm{BH}}-\sigma$ relation is different from that obtained via the
\citet{HarRix04} $M_{\rm{BH}}-M_{\rm{Bulge}}$ relation. The former peaks at a higher mass, by $\approx 0.1-0.2$
dex, and indicates a smaller number of black holes at the low mass end and a larger number of black holes
at the high mass end \citep[see also][]{LauFabRic07}.

\section{Discussion}
\label{sec:dis}

\subsection{Comparison with previous work}
\label{sec:dis1}

Using published distribution functions of the velocity dispersion in early and late-type galaxies
\citep[from][]{SheBerSch03}, and
the \citet{TreGebBen02} $M_{\rm{BH}}-\sigma$ relation, \citet[][see also \citealt{MarRisGil04}]{ShaSalGra04}
found that 29 per cent of the mass in black holes
is in late-type galaxies. The separation between early and late-type galaxies was done as in
\citet{BerSheAnn03a}, i.e. mainly with a cut in the concentration index $R90/R50$ at 2.5.
\citet{GraDriAll07} used the $M_{\rm{BH}}-n$ relation (where $n$ is the bulge S\'ersic index),
and luminosity distribution functions for red and blue spheroids \citep[from][]{DriAllLis07}, and
found a corresponding fraction of 22 per cent. To directly compare these results with ours is difficult
due to the fact that early-type (or red) galaxies in such studies should include not only ellipticals but
also a substantial fraction of galaxies with classical bulges, and some pseudo-bulges as well.
While 99 per cent of our ellipticals
have $R90/R50>2.5$, this is also the case for 76 per cent of our galaxies with classical bulges, and 8
per cent of our galaxies with pseudo-bulges.
\citet{BerSheAnn03a} also used other criteria to separate
early-type galaxies, but it is unlikely that these criteria excluded most disc galaxies.
Thus, the fact that we find that 45 per cent of the mass in black holes is in bulges, a higher fraction than
the previous estimates for late-type galaxies, is naturally expected. However, we can assume that the
``early-type galaxy'' bin in these previous studies includes the same fractions of elliptical galaxies and
galaxies with classical and pseudo-bulges as those we find in our sample using the same threshold in
concentration. Since this is the main morphological
criterion applied to define the samples used
by \citet{SheBerSch03}, it should allow a rough comparison at least with the results from
\citet{ShaSalGra04}. We thus combine 99 per cent of our estimate of
the black hole mass in ellipticals, 76 per cent of that in classical bulges, and 8 per cent of that in
pseudo-bulges, to obtain a total
black hole mass that can be compared to that in the early-type galaxies of \citet{ShaSalGra04}.
Applying these zeroth-order corrections, we get that 14 per cent of the mass
in black holes is in what could be called late-type galaxies, already accounting for our selection
effect due to the axial ratio cut. This is below the estimates in both \citet{ShaSalGra04} and
\citet{GraDriAll07}, but consistent with that in \citet{GraDriAll07}, with the uncertainties they quote.
Most likely, our estimate is below these previous results due to our cut in total stellar mass
at the low mass end. \citet{ShaSalGra04} and \citet{GraDriAll07} include galaxies with masses below
our mass cut, and thus a significantly larger fraction of late-type galaxies, as the fraction of
late-type galaxies is strongly increasing as one goes lower in mass.
The larger difference with respect to \citet{ShaSalGra04} might be at least partially a result from the
different relations used to estimate $M_{\rm{BH}}$.

\citet{MalBerBla09} also investigate quantitatively how late-type galaxies can be misclassified as
early-type galaxies, due to dust reddening alone, if one applies a colour cut in order to do such
a classification in the SDSS \citep[see also][]{MitKeeFri05}.
They find that the ratio of red to blue galaxies changes from 1:1 to 1:2 when going from observed to
intrinsic colours. Therefore, the true fraction of disc galaxies rises by a factor of about 1.3. This results
only from inclined disc galaxies being misclassified as ellipticals because dust attenuation makes
their colours too red for a typical disc galaxy. Our estimate of the fraction of the total black hole mass
in bulges is a factor of 1.5 higher than that in \citet{ShaSalGra04}, and a factor of 2 higher than
that in \citet{GraDriAll07}. Thus, the effects of dust reddening alone cannot explain this difference.
However, we argue that an important fraction of intrinsically red and concentrated lenticulars and early-type
spirals (with massive bulges and black holes) are present in the early-type/red samples in both studies mentioned,
and thus the masses of their black holes are being computed together with black holes in ellipticals.
Since lenticulars and early-type spirals generally have a relatively low dust content, this effect is
not being account for in the estimate of \citet{MalBerBla09}. Hence, the larger fraction of the total
black hole mass we find in bulges is not an unexpected result.

\subsection{Bulge formation and black hole growth}
\label{sec:dis2}

The results described above indicate that pseudo-bulges have higher $\sigma$ for their masses,
as compared to classical bulges, which also have, on average, higher $\sigma$ for their masses
when compared to elliptical galaxies.
We have kept the low $\sigma$ estimates from SDSS
for pseudo-bulges to avoid artificially strengthening these results.
For these systems, it is unclear, however,
if such results arise from the inability of SDSS measurements to
correctly measure $\sigma$ in cases where the true velocity dispersion is lower than, or close
to, the instrumental resolution. In other words, one could argue that the pseudo-bulges
for which $\sigma$ is available are only those at the high end of the velocity dispersion
distribution. As mentioned above, we have $\sigma$ estimates for only 30 per cent of the pseudo-bulges
in our sample. In such a case, our results would only indicate that pseudo-bulges display a very large
scatter around the $M_{\rm{Bulge}}-\sigma$ relation. While we can not presently rule out
such possibility, we note that there seems to be a gradual transition from elliptical
galaxies to classical bulges and pseudo-bulges in Fig. \ref{fig:smbh} and in the edge-on view
of the fundamental plane (Paper I).
This suggests that our results are the outcome of a real effect,
since classical bulges have $\sigma$ estimates typically substantially larger than the
SDSS instrumental resolution.

The result that pseudo-bulges have higher velocity dispersion than classical bulges,
at a fixed stellar mass, goes in the opposite direction as one would
naively infer from the virial theorem, particularly because pseudo-bulges are more
rotationally supported than classical bulges. However, pseudo-bulges are not expected to
be fully relaxed systems, which is one of the main assumptions in the virial theorem.
Concerning this result, one may worry that disc contamination in the SDSS fibre (from
which the spectra, and thus $\sigma$, are obtained) could artificially rise the values
of $\sigma$ in pseudo-bulges, where such contamination is expected to be present to
some degree. In fact, disc rotation can result in overestimated $\sigma$ values, since
with SDSS data alone one can not distinguish true dispersion from rotation.
This is also true for bulge rotation and, again, such effect is expected to be
more significant for pseudo-bulges than for classical bulges. However, given that
we have only selected galaxies with $b/a\geq0.9$, i.e. face-on galaxies, both disc
and bulge rotation should have negligible components along the line of sight, and
the effects from rotation are {\em not} expected to be present. Indeed, the intrinsic
axial ratio of discs seems to be closer to 0.9 rather than exactly 1
\citep[see e.g.][]{Ryd04,Ryd06}. Although recent numerical experiments yield
similar results \citep{YouHopCox08}, it remains to be verified if the different behaviour
of pseudo-bulges in the $M_{\rm{Bulge}}-\sigma$ relation is not only a consequence
of the presence of bars \citep[see also][]{Gra08}.
A bar is expected to enhance the central velocity dispersion in its
host galaxy {\em even in face-on galaxies}, if it has evolved
for a sufficient time \citep[see][and references therein]{GaddeS05},
and such effect should be more dramatic in galaxies with less conspicuous bulges.
It should also be noted that it is likely that a fraction of our unbarred galaxies
contain small bars (smaller than $2-3$ kpc in semi-major axis) which have been
missed due to the relatively poor spatial resolution of SDSS images. Such fraction should
be larger in galaxies with pseudo-bulges, since these bars are more often hosted by
galaxies with small bulge-to-total ratios. This could explain the few pseudo-bulges
in ``unbarred'' galaxies that are also outliers in the $M_{\rm{Bulge}}-\sigma$ relation, should
the exception caused by pseudo-bulges be only due to the presence of a bar.

\citet{BerSheTun07} have recently shown that the relation between elliptical/bulge {\em luminosity}
$L$ and $\sigma$ in the local sample of galaxies with black hole mass measurements appears to differ
from that in a sample of SDSS early-type galaxies. It is presently unclear if this is a selection or a
physical effect, but it has important consequences, since the
distributions of black hole masses estimated using SDSS data depends on whether luminosity or
$\sigma$ is used to derive $M_{\rm{BH}}$. The results in \citet{Gra08} indicate that once barred
galaxies are removed from the samples with measurements of $M_{\rm{BH}}$ the $L-\sigma$ relation
so obtained is consistent with that in the SDSS sample. It also seems that one should not consider
all bulges together in these analyses \citep[as suggested in][]{Hu08}, although again such discrepancy
between classical and pseudo-bulges is likely to be related to the presence of a bar, as argued
in \citet{Gra08}, and confirmed in this work. In fact, we have shown here that the observed
$M_{\rm{Bulge}}-\sigma$ relation is different for ellipticals, classical bulges and pseudo-bulges,
a result which most likely originates from different formation and evolutionary histories.
A consequence from this result is that either the $M_{\rm{BH}}-M_{\rm{Bulge}}$ relation or the
$M_{\rm{BH}}-\sigma$ relation (or even both), has to have different forms for ellipticals
and classical bulges, and also possibly for pseudo-bulges. If so, then one may not need to invoke
distinct black hole fuelling mechanisms to explain different $M_{\rm{BH}}-M_{\rm{Bulge}}$
or $M_{\rm{BH}}-\sigma$ relations for different stellar systems. The existence of such different
relations is yet another important detail that has to be taken into account by studies on black hole
demographics and the connection between the properties of black holes and their
host galaxies.

\section*{Acknowledgments}
We are grateful to Peter Erwin, Tim Heckman and Simon White for useful discussions, and an
anonymous referee for comments that helped improved the paper. We thank Eyal Neistein
for his help in convolving the distribution functions.
DAG is supported by the Deutsche Forschungsgemeinschaft and the Max Planck Society. Funding
for the SDSS project has been provided by the Alfred P. Sloan Foundation,
the SDSS member institutions, the National Aeronautics and Space
Administration, the National Science Foundation, the US Department
of Energy, the Japanese Monbukagakusho, and the Max Planck
Society.
\bibliographystyle{mn2e}
\bibliography{../../gadotti_refs}

\label{lastpage}

\end{document}